\documentclass[aps,superscriptaddress,nofootinbib,eqsecnum,prd,notitlepage,twocolumn]{revtex4-1} 

\pdfoutput=1

\usepackage{amsfonts}
\usepackage{amsmath}
\usepackage{amssymb}
\usepackage{graphicx,color}
\usepackage{float}
\usepackage{hyperref}
\usepackage{subfigure}
\usepackage{dcolumn}
\usepackage{soul}
\usepackage{ulem}
\usepackage{verbatim}

\begin{document}

\title{Graceful exit problem in warm inflation}

\author{Suratna Das}
\email{suratna@iitk.ac.in}
\affiliation{Department of Physics, Indian Institute of Technology,
   Kanpur 208016, India}

\author{Rudnei O. Ramos}
\email{rudnei@uerj.br}
\affiliation{Departamento de Fisica Teorica, Universidade do Estado
  do Rio de Janeiro, 20550-013 Rio de Janeiro, RJ, Brazil }

\begin{abstract}

A seemingly simple question, how does warm inflation exit gracefully?,
has a more complex answer than in a cold paradigm. It has been
highlighted here that whether warm inflation exits gracefully depends
on three independent choices: The form of the potential, the choice of
the warm inflation model (i.e., on the form of its dissipative
coefficient) and the regime, of weak or strong dissipation,
characterizing the warm inflation dynamics. Generic conditions on
slow-roll parameters and several constraints on the different model
parameters required for warm inflation to exit gracefully are derived. 

\end{abstract}

\maketitle

\section{Introduction}
\label{intro}

Warm inflation (WI)~\cite{Berera:1995ie}, a variant inflationary
paradigm, where the inflaton field dissipates its energy to a thermal
bath throughout inflation, has many attracting features than its
counterpart, the cold inflationary paradigm. First of all, WI takes
into account the interactions of the inflaton field with other
particle degrees of freedom during inflation, which are in general
ignored or considered negligible in a cold inflation paradigm.  Due to
these interactions of the inflaton field, when WI ends, it ends in a
universe which has an existing thermal bath. This is unlike the case
of the cold paradigm where the universe ends up in a state devoid of
matter at the end of inflation, thus requiring a subsequent phase of
(p)reheating.  As WI smoothly ends in a radiation dominated universe,
it avoids a following reheating phase --- physics of which is yet not
fully understood.  Secondly, the dissipative effects that are
intrinsic to the WI picture offers an opportunity of alleviating  some
of the long-lasting problems related to cold inflation. In particular,
recent works on WI have shown that for sufficiently strong
dissipation, sub-Planckian inflaton values are allowed during
inflation~\cite{Bastero-Gil:2019gao,Kamali:2019xnt,Das:2020xmh}.  This
supports the current data~\cite{Akrami:2018odb} which prefer small
field models over the large-field ones.  It also leads to a lower
radiation temperature that significantly alleviates and possibly
solves issues related to overproduction of unwanted relics and, e.g.,
the gravitino problem~\cite{Sanchez:2010vj,Bartrum:2012tg}.  WI has
also been shown (see, e.g., Ref.~\cite{Bastero-Gil:2019gao}) to
provide a natural solution to the so-called eta-problem plaguing some
cold inflation models in the supergravity context.  In addition, it
has been shown that  an appropriate baryon asymmetry can possibly be
generated by dissipative effects alone during warm
inflation~\cite{BasteroGil:2011cx}. This, on the other hand, yields
additional observable baryon isocurvature
perturbations~\cite{Bastero-Gil:2014oga}, which can be probed to check
the  consistency of WI models.  Third, WI  has a more enhanced scalar
curvature perturbation spectrum, which yields a lesser
tensor-to-scalar ratio and, thus, helps accommodate  models that would
otherwise be ruled out by the observations in a cold
paradigm~\cite{Akrami:2018odb}. The monomial chaotic inflaton
potentials~\cite{Benetti:2016jhf,Berera:2018tfc} are one such example.
Fourth, but not the least, it has been shown that the
fluctuation-dissipation dynamics, which is an inherent feature of WI
model realizations, can help solving the initial condition problem of
plateau-like potentials and at the same time can  work as a mechanism
which helps localize the inflaton at the origin to trigger a period of
sufficient slow-roll inflation~\cite{Bastero-Gil:2016mrl}.

In its initial days, WI was considered hard to be implemented through
a consistent microscopic model, with the difficulty been mostly
related to the problem of shielding the inflaton sector from large
thermal corrections~\cite{Berera:1998gx,Yokoyama:1998ju}, which would
endanger the required flatness of the inflaton potential. However,
this difficulty was soon overcome, first by decoupling the inflaton
sector from the light radiation fields (for its model building
construction, see Refs.~\cite{Berera:2008ar,BasteroGil:2012cm}, while
for the consistency with observations, see
Ref.~\cite{Bartrum:2013fia}) and, later, by  using an analogous model
motivated by the Higgs phenomenology, it was shown how the inflaton
could be coupled directly to the light radiation fields, yet
preventing any harmful large thermal corrections to the inflaton
potential. This was initially shown to be possible when the inflaton
was directly coupled to light fermion
fields~\cite{Bastero-Gil:2016qru}, a model that was named  Warm Little
Inflaton (WLI) model, and more recently through a variant of the WLI,
where the inflaton was coupled directly to light scalar bosonic
fields~\cite{Bastero-Gil:2019gao}. We will call the latter the variant
of Warm Little Inflaton (VWLI) model. Another model with similar
features to the WLI and VWLI, but motivated by the physics  of axions
and natural inflation has been named the Minimal Warm Inflation (MWI)
model~\cite{Berghaus:2019whh}. One thing which is common to all these
models is that the dissipation coefficient typical of the WI dynamics
can all be expressed through a simple functional form that is
dependent on the temperature $T$ of the thermal bath and the amplitude
$\phi$ of the inflaton field.  The simplest functional form assumed in
most phenomenological studies involving WI involves a dissipation
coefficient $\Upsilon$ given by
\begin{equation}
\Upsilon(\phi,T)=C_\Upsilon\,T^p \phi^c M^{1-p-c},
\label{Upsilon}
\end{equation}
where $C_\Upsilon$ is a dimensionless constant (that carries the
details of the microscopic model used to derive the dissipation
coefficient, e.g., the different coupling constants of the model ---
see, for example, Ref.~\cite{BasteroGil:2010pb} for different
$C_\Upsilon$ expected for the dissipation coefficient in WI, depending
on the interactions involved and regime of parameters). The numerical
powers given by $c$ and $p$, can be either positive or negative
numbers and $M$ is some appropriate mass scale, such that the
dimensionality of the dissipation coefficient in Eq.~(\ref{Upsilon})
is preserved, $[\Upsilon] = [{\rm energy}]$.

WI has also more recently regained attention in the literature in the
context of the so-called Swampland
Conjectures~\cite{Obied:2018sgi,Garg:2018reu,Ooguri:2018wrx,Kinney:2018nny}.
The  Swampland Conjectures have been formulated as conditions that an
effective field theory should satisfy in order to accommodate an
ultraviolet complete field theory within a String landscape. {}For
instance, in terms of the standard slow-roll coefficients,
\begin{eqnarray}
\epsilon_V=\frac{M_{\rm Pl}^2}{2}\left(\frac{V'}{V}\right)^2,\quad
\eta_V=M_{\rm Pl}^2\frac{V''}{V},
\label{epseta}
\end{eqnarray}
with $M_{\rm Pl} \equiv 1/\sqrt{8 \pi G} \simeq 2.4 \times 10^{18}$GeV
is the reduced Planck mass, the de Sitter conjecture, requires either
$\epsilon_V>\mathcal{O}(1)$ or $\eta_V<-\mathcal{O}(1)$
\cite{Garg:2018reu,Ooguri:2018wrx}. Thus, the de Sitter swampland
conjecture alone tends to overrule the generic conditions for
inflation, which requires $\epsilon_V\ll1$, $\eta_V\ll1$ instead. This
makes it difficult for the usual cold inflation to be realized in the
landscape of a string theory. As WI modifies the slow-roll conditions,
which are now expressed as $\epsilon_V\ll1+Q$ and $\eta_V\ll1+Q$,
where the dimensionless quantity $Q \equiv \Upsilon/(3H)$ can be
larger than 1,  this paradigm can much easily accommodate the
Swampland Conjectures.  This was first noted in
Ref.~\cite{Das:2018hqy}, and later, quite a few discussions along the
same line have
followed~\cite{Motaharfar:2018zyb,Das:2018rpg,Bastero-Gil:2018yen,Bastero-Gil:2019gao,Kamali:2019xnt,Das:2019hto,Berera:2019zdd,Goswami:2019ehb,Brandenberger:2020oav,Berera:2020dvn,Das:2020xmh}. 

Aside its many attractive features, ending inflation in WI scenario
is, in a sense, a more complex process than in the cold paradigm, and
this is the main aim of this article  --- analyze how (and whether) WI
gracefully exits. In a generic cold paradigm, inflation takes place
when $\epsilon_H\sim\epsilon_V\ll1$ and terminates when  $\epsilon_V$
becomes of the order 1 (here $\epsilon_H$ and $\epsilon_V$ are the
Hubble  and potential slow-roll parameters in their conventional
forms, respectively). Hence, the cold inflation paradigm is always in
need of  appropriate forms of potentials which yield growing slow-roll
parameters ($\epsilon_V$) in time (or with number of $e$-foldings
$N$). On the other hand, in the WI paradigm, inflation ends when
$\epsilon_V\sim1+Q$. This comes from the fact that, in WI, the Hubble
slow-roll parameter becomes $\epsilon_H\sim \epsilon_V/(1+Q)$
\cite{Berera:2008ar,BasteroGil:2009ec}, and thus inflation ends when
the Hubble slow-roll parameter becomes of the order unity
($\epsilon_H\sim1$) or, equivalently, the potential slow-roll
parameter, $\epsilon_V$, becomes of the order $1+Q$. However, $Q$, in
general, evolves during inflation, as well as the potential slow-roll
parameter $\epsilon_V$. Both can increase, decrease or even can remain
constant depending on the form of the potential, the region (strong
($Q\gg1$) or weak ($Q\ll1$)) in which WI is taking place and also on
the form of the dissipative coefficient (i.e., on the choice of the WI
model). Note also that the end of the slow-roll inflationary phase in
WI is  also related by the fact that radiation can smoothly overtake
the inflaton potential energy density at end. Putting in these terms,
the problem of graceful exit can also be expressed by the fact that
even though radiation is being produced  as a consequence of
dissipation, it might never overtake the inflaton energy density.
Without an efficient mechanism to shut down dissipation, inflation can
then prolong to asymptotic large times, with both radiation and
inflaton energy densities decreasing but without radiation becoming
dominant. We will give explict examples where this can happen. Hence,
graceful exit is a more complex issue in WI than it is in cold
inflation. 

To our knowledge, a systematic study of the relevant parameter space
corresponding to typical WI models that fully leads to a graceful exit
is still lacking in the literature. The main objective of this work is
exactly to fill this important gap. Besides of investigating this
issue, we also analyze, as a consequence of our study, those generic
WI potential and  dissipation model parameters region of space where
the dissipation ratio can grow or decrease. This is of particular
importance for example in several model buildings in WI.  

In the next section, Section~\ref{sec2}, we introduce the problem in
more details.  Then, in Section~\ref{sec3}, we discuss our results
when considering different forms of  dissipation coefficients
typically considered in many WI studies and also with different large
classes of inflaton primordial potentials. Potential applications of
our results are discussed in Section~\ref{applications}. {}Finally, in
Section~\ref{conclusions}, we give our conclusions and final remarks.

\section{Setting up the problem}
\label{sec2}

To look into the matter more closely, we need to first look at the
basic background dynamics of WI. In the leading adiabatic
approximation, the dynamics of the inflaton field in WI has an extra
friction term arising due to dissipation,
\begin{eqnarray}
\ddot\phi+3H(1+Q)\dot\phi=-V'(\phi),
\label{eqphi}
\end{eqnarray}
while the dynamics of the radiation energy density $\rho_R$ is given
by
\begin{equation}
\dot \rho_R + 4 H \rho_R = 3H Q \dot \phi^2,
\label{eqrhoR}
\end{equation}
where dots denote temporal derivatives, $H$ is the Hubble parameter,
\begin{equation}
H^2 \equiv \left(\frac{\dot a}{a}\right)^2 =\frac{1}{3 M_{\rm Pl}^2}
\left( \frac{\dot \phi^2}{2} + V + \rho_R \right),
\label{Hubble}
\end{equation}
where $a$ is the scale factor and the dimensionless quantity $Q$ is
defined as,
\begin{eqnarray}
Q\equiv \frac{\Upsilon}{3H},
\end{eqnarray}
$\Upsilon$ being the dissipation coefficient. In general, $\Upsilon$
can be a function of both $\phi$ and $T$. Different forms of
dissipative coefficients, derivable from nonequilibrium quantum field
theory methods, have been studied extensively in the
literature~\cite{Berera:1998gx,Berera:2008ar,BasteroGil:2009ec,Zhang:2009ge,BasteroGil:2010pb,BasteroGil:2012cm}.

The graceful exit problem can be simply formulated as
follows. Inflation takes place in the WI scenario when
$\epsilon_V<1+Q$ and ends when $\epsilon_V\sim 1+Q$.  Therefore, if
$Q$ increases with the number of $e$-foldings, then $\epsilon_V$ has
to increase faster than $Q$ in order to end inflation. On the other
hand, when $Q$ decreases, inflation naturally ends if $\epsilon_V$
increases with the number of $e$-foldings or just remains as a
constant. Otherwise, $\epsilon_V$ has to decrease with a slower rate
than $Q$ in order to end inflation. To put it simply, we have that
\begin{eqnarray}
\epsilon_H\equiv -\frac{\dot H}{H^2} \simeq \epsilon_{wi}\equiv
\frac{\epsilon_V}{1+Q},
\label{epsH}
\end{eqnarray}
which implies that inflation takes place when $\epsilon_{wi}<1$ and
ends when $\epsilon_{wi}\sim 1$.\footnote{The condition
  $\epsilon_{wi}=1$, in the true sense,  is a weaker condition to end
  WI than $\epsilon_H=1$, as the first condition suggests that  WI
  ends when $\rho_R=V/2$ (when $Q\gg1$), whereas, in reality, WI tends
  to end when  the radiation energy density equals and surpasses the
  potential energy density. Therefore, the weaker condition,
  $\epsilon_{wi}=1$ predicts the end of inflation slightly
  earlier. However, it is to note that, in all practical purposes,
  the weaker condition only underestimates the end of inflation by
  less than one $e$-folding.  Therefore, using the stronger condition
  instead of the weaker one would not alter the bounds we  would be
  obtaining at the later part of the paper.} This shows that
$\epsilon_{wi}$ has to grow with the number of $e$-foldings $N$ in
order to end inflation, yielding the condition 
\begin{eqnarray}
\frac{d\ln\epsilon_V}{dN}>\frac{Q}{(1+Q)}\frac{d\ln Q}{dN},
\label{end-inf-cond}
\end{eqnarray}
where $dN=Hdt$.
It is now quite apparent from Eq.~(\ref{end-inf-cond}) that if $Q$ is
an increasing function of $N$, then $\epsilon_V$ has to increase
faster than that (noticing the fact that both $\epsilon_V$ and $1+Q$
are positive quantities). On the other hand, when $Q$ decreases with
$N$, then either $\epsilon_V$ can grow at any rate or may not evolve
at all. But if $\epsilon_V$ decreases too, then it has to decrease
slower than $Q$. All these conditions have been stated in the previous
paragraph. It is also clear from Eq.~(\ref{end-inf-cond}) that when
$Q$ does not evolve with $e$-foldings, $\epsilon_V$ can grow at any
rate, just like in the cold inflation scenario, in order to end
inflation. 

It is to note here that in cold inflation, inflation generally does
not end when the inflaton field gets trapped into some false
vacuum. However, that is not necessarily the only reason behind a
non-graceful exit in the case of WI. We can see from the above
discussion that, mostly, inflation does not end in cases in Warm
inflation when $Q$ grows faster than $\epsilon_V$. As the presence of
$Q$ in the equation of motion of the inflaton field acts like an extra
frictional term over the one for the expansion term, an increasing $Q$
will eventually lead to an overdamped equation of motion for the
inflaton field. That means, that the inflaton field will only reach
the minimum of the potential in asymptotic time, resulting in a
scenario of never-ending WI.

We will now derive the general conditions under which WI undergoes a
graceful exit. In order to do so,  we assume a generic parametrization
for the dissipation coefficient $\Upsilon$ as a function of $\phi$ and
the temperature $T$ as given by Eq.~(\ref{Upsilon}). The
parametrization  given by Eq.~(\ref{Upsilon}) covers a large class of
WI models that have been studied before in the literature. {}For
instance, early dissipation coefficient derived in
Refs.~\cite{Gleiser:1993ea,Berera:1998gx}, corresponds to $p=-1,c=2$,
the one derived in Refs.~\cite{Berera:2008ar,BasteroGil:2012cm}
corresponds to $p=3,c=-2$. The dissipative coefficient, $\Upsilon$, in
WLI model varies linearly with the temperature $T$, as
$\Upsilon(T)\propto T$ \cite{Bastero-Gil:2016qru}, corresponding to
$p=1,c=0$ in Eq.~(\ref{Upsilon}), whereas in MWI model it varies with
a cubic power of $T$, $\Upsilon(T)\propto T^3$
\cite{Berghaus:2019whh}, where $p=3,c=0$. However, the temperature
dependence of $\Upsilon$ in VWLI model is more complex, given
by~\cite{Bastero-Gil:2019gao}
\begin{equation} \label{dissip}
\Upsilon\simeq {C}{M^2 T^2\over
  m^3(T)}\left[1+{1\over\sqrt{2\pi}}\left({m(T)\over
    T}\right)^{3/2}\right]e^{-m(T)/T},
\end{equation}
where $M$ is a mass scale of the model, $m^2(T) = m_0^2 + \alpha^2
T^2$ is the thermal mass for the light scalars coupled to the inflaton
in the model of VWLI, with $m_0$ the vacuum mass of those light
scalars and $\alpha$ is a coupling constant (actually, a combination
of coupling constants appearing in the Lagrangian density of the
model).  Considering the leading behavior when the effective mass is
dominated by its thermal part, $m(T) \sim \alpha T$, the dissipation
coefficient (\ref{dissip}) varies as $\Upsilon(T)\propto T^{-1}$,
realizing the case with $p=-1,c=0$ in Eq.~(\ref{Upsilon}).  On the
other hand, as the temperature of the thermal bath drops and the
vacuum term $m_0$ in  $m(T)$ starts no longer to be negligible, it
effectively would correspond to values of $p > -1$, with a limiting
case of $p=2$ when $m_0 \gg \alpha T$ (with an exponentially
suppressed dissipation).  The parametrization assumed here for the
dissipation coefficient, Eq.~(\ref{Upsilon}), also covers other
particular forms for the dissipation coefficient and used in earlier
phenomenological studies, like the case $p=0,\,c=0$ (a constant
dissipation coefficient form), which can be considered as a particular
case that can emerge for some specific dynamical regimes in the more
general dissipation  coefficient Eq.~(\ref{dissip}).  Also, cases with
$p=0$ (but with particular powers of the inflaton field), e.g.,  such
that $\Upsilon \propto H$, can mimic many previous phenomenological
studies of WI dynamics, since it leads strictly to a constant
dissipation ratio $Q=\Upsilon/(3H)$. Such a form leading to a constant
dissipation ratio is particularly useful for deriving analytical
results in WI and has been employed by many authors before exactly
because of that (see also, e.g.,
Refs.~\cite{Zhang:2009ge,Herrera:2013rra,Herrera:2014mca,Visinelli:2016rhn,Jawad:2017rkq,Jawad:2017gwa}
for different applications in WI using a generic form for the
dissipation coefficient).  

In the analysis that follows now, we will also employ the slow-roll
approximated dynamical equations of WI, which follow directly from
Eqs.~(\ref{eqphi}), (\ref{eqrhoR}) and (\ref{Hubble}),
\begin{eqnarray}
\!\!\!\!\!\!\! 3H(1+Q)\dot\phi\simeq -V',\quad \rho_R \simeq \frac{3
  Q\dot\phi^2}{4},\quad 3H^2 \simeq\frac{V}{M_{\rm Pl}^2}.
\label{slowrolleqs}
\end{eqnarray}
It is important to note here that the slow-roll conditions in WI do
not imply the slow-roll parameters $\epsilon_V$ and $\eta_V$ to be
smaller than unity. One can have these slow-roll parameters to be
large, while still maintaining the required conditions for inflation,
e.g., $\epsilon_H \ll 1$, provided that we have  $Q \gg 1$,  as one
can see from Eq.~(\ref{epsH}).  {}From the previous equations, one can
find out, after simple manipulations, that $Q$ is related to the form
of the chosen potential as 
\begin{eqnarray}
(1+Q)^{2p}Q^{4-p}&=&\frac{M_{\rm
      Pl}^{2p+4}C_\Upsilon^4M^{4(1-p-c)}}{2^{2p}
    3^2\,C_R^p}\frac{V'^{2p}}{V^{p+2}}\phi^{4c}\nonumber\\ & \equiv&
  \tilde C \frac{V'^{2p}}{V^{p+2}}\phi^{4c},
\end{eqnarray}
where $C_R$ is the Stefan-Boltzmann factor relating the radiation
energy density to the temperature, $\rho_R = C_R T^4$. This allows us
to determine how $Q$ evolves during WI,
\begin{eqnarray}
C_Q\frac{d\ln Q}{dN}=(2p+4)\,\epsilon_V-2p\,\eta_V-4c\, \kappa_V,
\label{q-evo}
\end{eqnarray}
where
\begin{eqnarray}
C_Q\equiv 4-p+(4+p)Q.
\label{cq}
\end{eqnarray}
In Eq.~(\ref{q-evo}), besides the standard slow-roll
coefficients~(\ref{epseta}), we have also defined 
\begin{eqnarray}
\kappa_V\equiv M_{\rm Pl}^2\frac{V'}{\phi V}.
\end{eqnarray}
We also note that $C_Q$ in Eq.~(\ref{cq}) is a positive quantity,
otherwise $C_Q<0$ leads to a condition $Q<(p-4)/(p+4)$. As, so far,
all the viable models of WI have temperature dependent dissipative
coefficients with $-4<p<4$, which is required from the stability
studies of the WI
dynamics~\cite{Moss:2008yb,delCampo:2010by,BasteroGil:2012zr}.  Thus,
we will treat $C_Q$ as a positive quantity henceforth. On the other
hand, depending on the form of the inflaton potential, $\epsilon_V$
evolves during WI as 
\begin{eqnarray}
\frac{d\ln\epsilon_V}{dN}=\frac{4\epsilon_V-2\eta_V}{1+Q}.
\label{epsilon-evo}
\end{eqnarray}

Eqs.~(\ref{end-inf-cond}), (\ref{q-evo}) and (\ref{epsilon-evo}),
along with the forms of the primordial inflaton potential and
dissipation coefficient in WI, give all we need for our analysis. 

\section{Determining the region of parameters for graceful exit in WI}
\label{sec3}

We will now discuss how WI makes a graceful exit on a case-by-case
basis.

\subsection{Increasing $\epsilon_V$}
\label{sec3A}

We see from Eq.~(\ref{epsilon-evo}) that $\epsilon_V$ increases for
potentials with $4\epsilon_V-2\eta_V>0$. This encompasses a large
class of inflaton potentials commonly used in inflationary theories,
such as the whole range of monomial chaotic potentials,
\begin{equation}
V(\phi) = \frac{V_0}{n} \left(\frac{\phi}{M_{\rm Pl}}\right)^n,
\label{Vpoly}
\end{equation} 
and the concave potentials, which are presently preferred by the data
and for which $\eta_V$ is negative~\cite{Akrami:2018odb}. Though the
conventional monomial potentials, namely the quadratic and the
quartic, are not in good agreement with the data in a cold
inflationary scenario~\cite{Akrami:2018odb}, they match with the
observations very well in a WI setup as they produce less
tensor-to-scalar ratio in such
cases~\cite{Benetti:2016jhf,Berera:2018tfc,Bartrum:2013fia}. Thus,
monomial potentials are still of importance as long as one is dealing
with WI. 

\subsubsection{Positive $\eta_V$}

{}First, let us consider the potentials for which $\eta_V$ is
positive, e.g., the monomial potentials, Eq.~(\ref{Vpoly}), which we
will take as an explicit example. Here, $\epsilon_V$ increases when
\begin{eqnarray}
\frac{\epsilon_V}{\eta_V}>\frac12.
\label{inc-eps}
\end{eqnarray} 
In such a case, $Q$ grows with $N$ if 
\begin{eqnarray}
n> \frac{2p-4c}{p-2},
\label{inc-Q}
\end{eqnarray}
with $n$ being the order of the monomial potential. Otherwise $Q$
would decrease or remains constant, and inflation ends without any
further requirement in those cases. {}For cases where $Q$ increases,
the inequality in Eq.~(\ref{end-inf-cond}) sets a condition on the
potential slow-roll parameters in order to end inflation as
\begin{eqnarray}
&&\frac{\epsilon_V}{\eta_V} > \frac{1}{2(4-p)+(6+p)Q}\left[ - 2c\,Q
    \frac{\kappa_V}{\eta_V} + 4 - p +4Q \right].\nonumber \\
\label{cond-end-inf-gen}
\end{eqnarray}

Let us first consider the class of WI models where the dissipative
coefficient is a functions of $T$ alone, i.e., cases with $c=0$ in
Eq.~(\ref{Upsilon}).  Then, the condition Eq.~(\ref{cond-end-inf-gen})
would read as 
\begin{eqnarray}
\frac{\epsilon_V}{\eta_V}>\frac{(4-p)+4Q}{2(4-p)+(6+p)Q}.
\label{cond-end-inf}
\end{eqnarray}
We note that the above condition depends primarily on whether WI is
taking place in a weak or in a strong dissipative regime. If inflation
is taking place in the weak regime, $Q\ll 1$, the
condition~(\ref{cond-end-inf}) yields 
\begin{eqnarray}
\left.\frac{\epsilon_V}{\eta_V}\right|_{\rm weak}>\frac12,
\label{weak-cond}
\end{eqnarray}
whereas if inflation is taking place in the strong dissipative regime,
$Q\gg 1$, then one gets
\begin{eqnarray}
\left.\frac{\epsilon_V}{\eta_V}\right|_{\rm strong}>\frac4{6+p}.
\label{strong-cond}
\end{eqnarray}
It is to note here that the conditions in Eqs.~(\ref{weak-cond}) and
(\ref{strong-cond}) are independent of that in
Eq.~(\ref{inc-eps}). The condition in Eq.~(\ref{inc-eps}) depends
solely on the choice of the inflaton potential, whereas then
conditions in Eqs.~(\ref{weak-cond}) and (\ref{strong-cond}) depend
also on the choice of the WI model, i.e., on the form of the
dissipative coefficient. However, the condition for having
$\epsilon_V$ evolving faster than $Q$ in the weak regime,
Eq.~(\ref{weak-cond}), is the same as that of having a growing
$\epsilon_V$, Eq.~(\ref{inc-eps}). This implies that if WI takes place
in a weak dissipative regime with a potential yielding a growing
$\epsilon_V$, WI will always end gracefully. However, that is not the
case when WI takes place in a strong dissipative regime. Yet, if the
condition in Eq.~(\ref{strong-cond}) is stronger than in
Eq.~(\ref{inc-eps}), i.e., $4/(6+p)>1/2$, which takes place when
$p<2$, then the condition in Eq.~(\ref{strong-cond}) determines
whether or not $\epsilon_V$ would increase faster than $Q$ to end
inflation. This, then, also implies that for models with $p\geq 2$, a
potential with growing $\epsilon_V$ is sufficient to end
inflation. Thus, in a strongly dissipative regime, the condition to
end inflation can be summed up as
\begin{eqnarray}
\left.\frac{\epsilon_V}{\eta_V}\right|_{\substack{\rm strong
    \\ p<2}}>\frac4{6+p}, \quad
\left.\frac{\epsilon_V}{\eta_V}\right|_{\substack{\rm strong \\ p\geq
    2}}>\frac12.
\label{strong-cond-1}
\end{eqnarray}

\begin{center}
\begin{figure}[!htb]
\subfigure[ Evolution case for $p=-1$, $c=0$ and
  $n=2$. ]{\includegraphics[width=7.5cm]{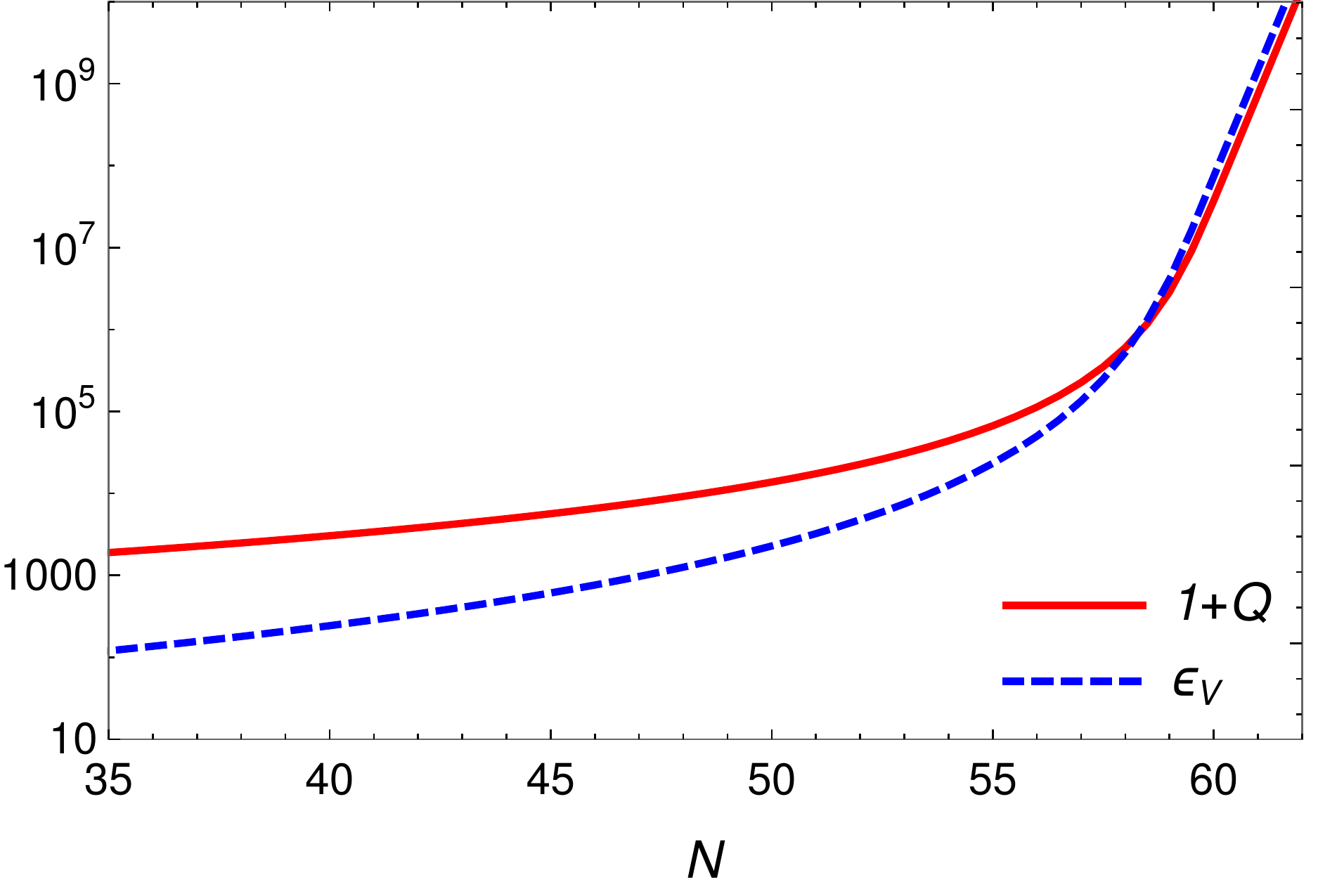}}
\subfigure[ Evolution case for $p=-1$, $c=0$ and
  $n=4$.]{\includegraphics[width=7.5cm]{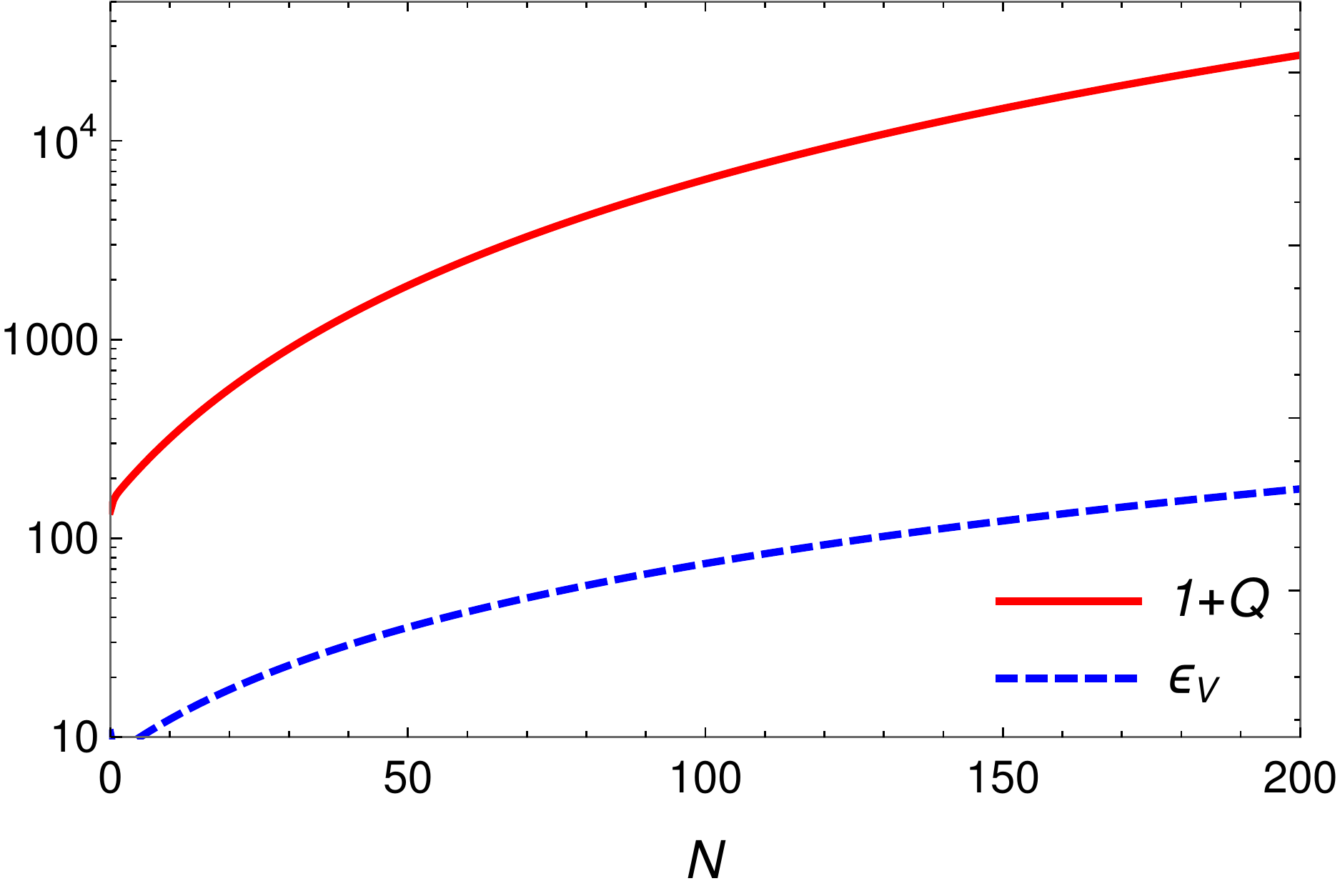}}
\subfigure[ Evolution case for $n=4$, but using
  Eq.~(\ref{dissip}). ]{\includegraphics[width=7.5cm]{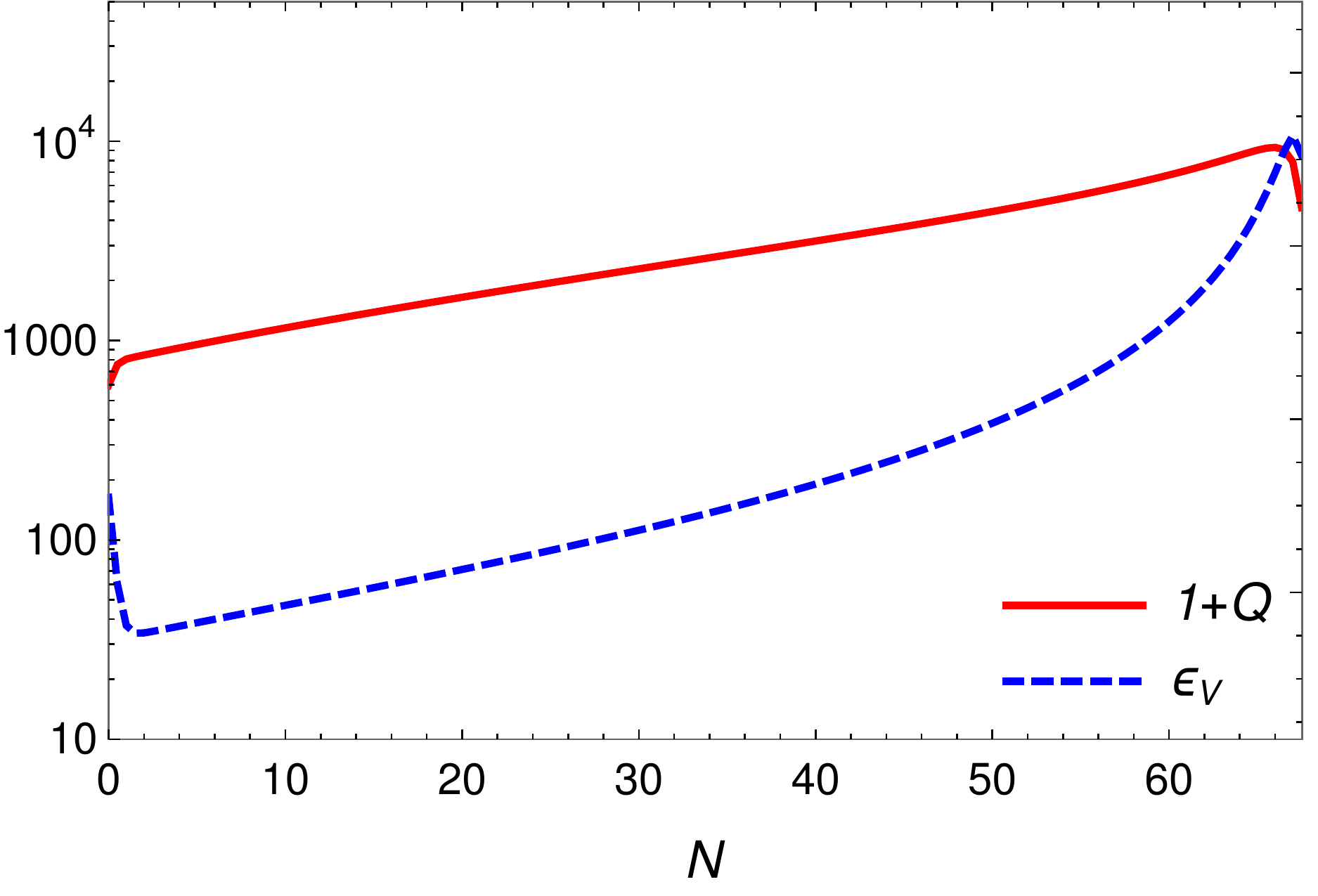}}
\caption{Example cases for the  evolution of $Q$ and $\epsilon_V$ with
  the number of $e$-foldings $N$. We have taken parameters analogous
  to those considered in Ref.~\cite{Bastero-Gil:2019gao} for
  convenience in this example.}
\label{fig1}
\end{figure}
\end{center}

\begin{center}
\begin{figure*}[!htb]
\subfigure[3d parameter space
  $(c,p,n)$]{\includegraphics[width=5.7cm]{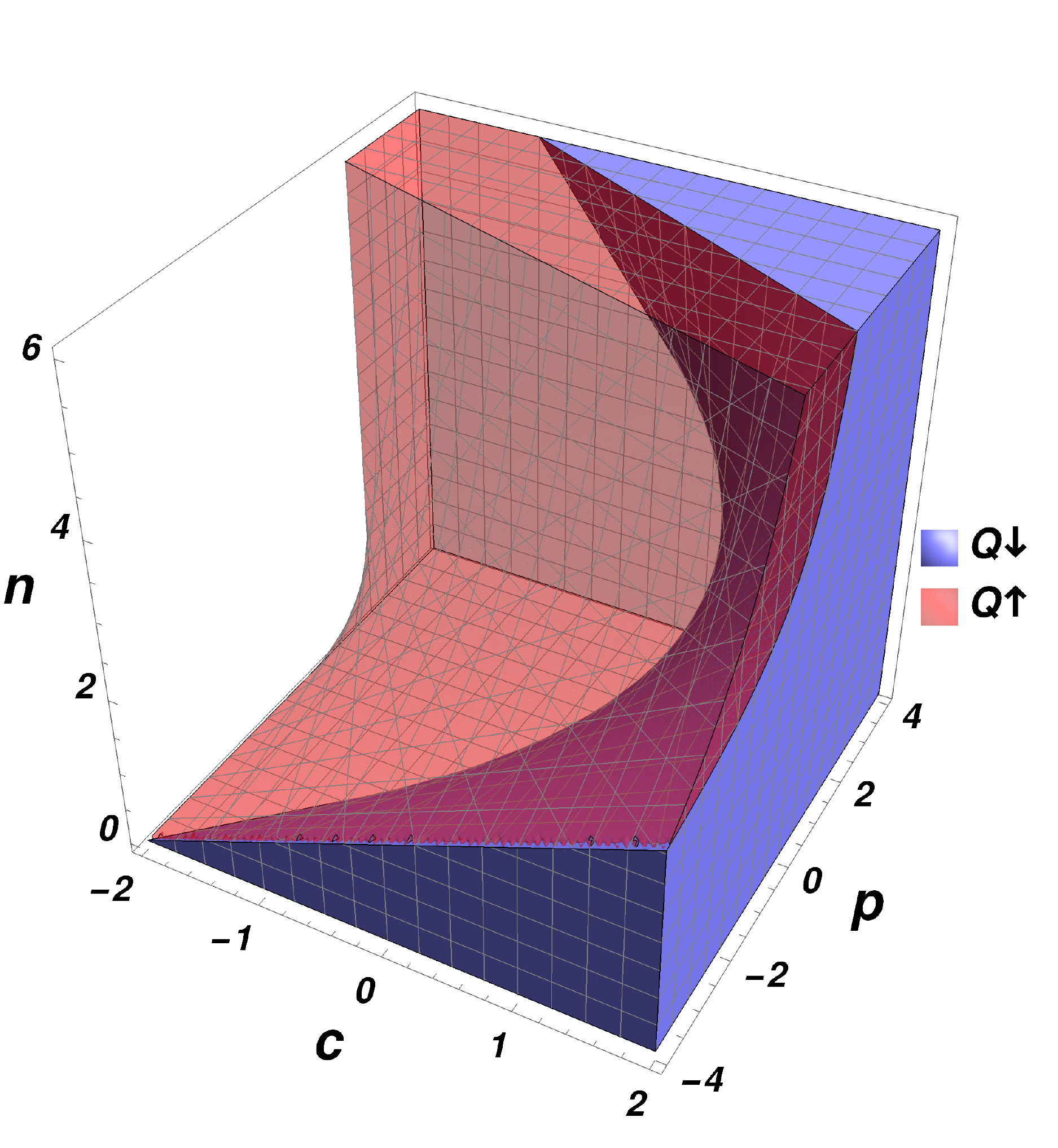}}
\subfigure[Plane $c=0$]{\includegraphics[width=5.2cm]{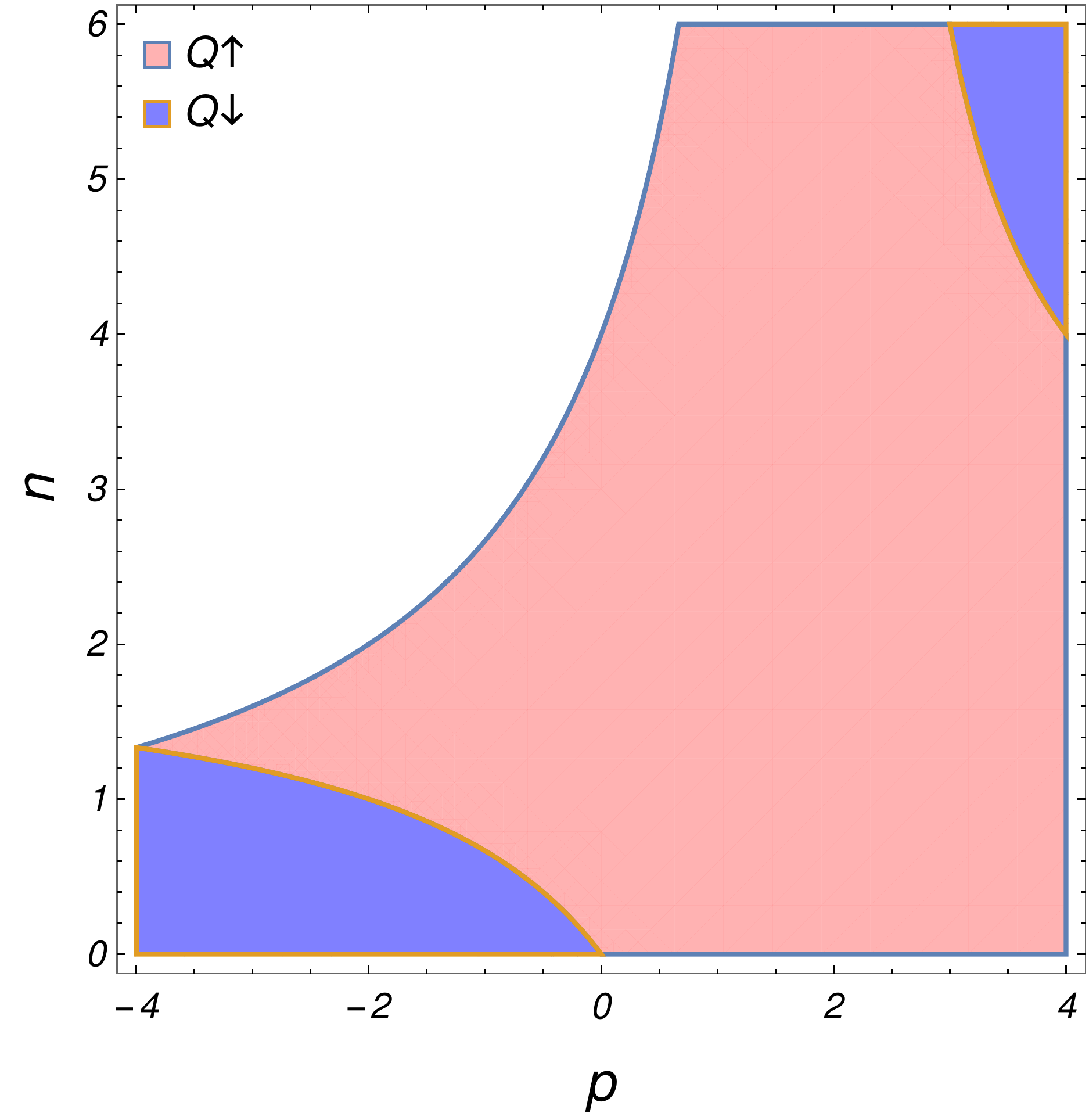}}
\subfigure[Plane $c=2$]{\includegraphics[width=5.2cm]{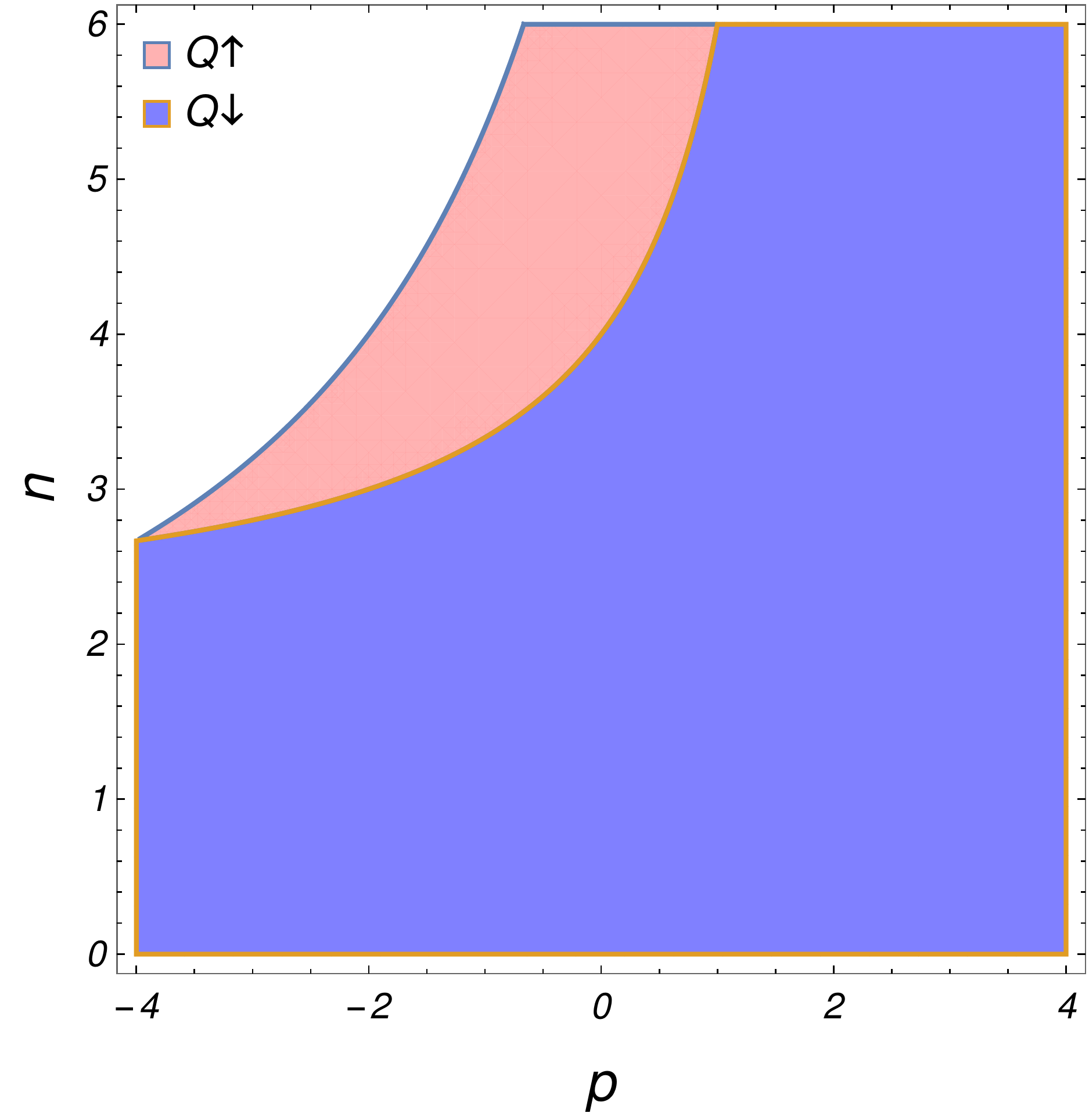}}
\caption{The  parameter space for graceful exit in the case of the
  monomial class of inflaton potentials. The shaded areas indicate
  where $Q$ grows or decreases and WI has a graceful  exit. Empty
  (white) space indicates where there is no graceful exit.}
\label{fig2}
\end{figure*}
\end{center}

The above discussion shows that except for models with $p<2$, where WI
is taking place in a strong dissipative regime, a form of potential
yielding growing $\epsilon_V$ is sufficient to end inflation, just
like in a cold inflation paradigm.  Moreover, for the generic form of
monomial potential Eq.~(\ref{Vpoly}), one gets a condition for
graceful exit for models with $p<2$ (while WI is taking place in the
strong dissipative regime) as 
\begin{eqnarray}
(2-p)n<8.
\end{eqnarray} 
Therefore, considering for example the case for the WLI
model~\cite{Bastero-Gil:2016qru}, with $p=1,c=0$, we obtain that $n<8$
in order to end inflation. This implies that the conventional monomial
potentials, like the quartic and the quadratic, help end inflation in
such a WI model.  However, for the VWLI
model~\cite{Bastero-Gil:2019gao}, when restricting to the case where
the thermal mass $m(T)$ in Eq.~(\ref{dissip}) is dominated by the
temperature dependent term, then the dissipative coefficient would
effectively evolve as inverse of the temperature, yielding $p=-1,c=0$
as already explained before. In such a case, inflation only ends with
a choice of a monomial potential with $n<8/3$ (while for $n<2/3$, $Q$
decreases,  as can be seen from Eq.~(\ref{inc-Q}) and inflation ends
without any further  requirement). This is an important example case
in WI, which demonstrates the importance of not only considering the
phenomenological form of the dissipation coefficient but also the
associated microscopic physics involved in the derivation of the
model. To show this explicitly, in {}Fig.~\ref{fig1} we show the
results for $\epsilon_V$ and $Q$ when evolving the background
evolution equations, given by Eqs.~(\ref{eqphi}), (\ref{eqrhoR}) and
(\ref{Hubble}), and for three cases: {}For the dissipation coefficient
Eq.~(\ref{Upsilon}) when taking $p=-1,c=0$ and in the case of a
quadratic potential $n=2$ (panel a), for $p=-1,c=0$ and in the case of
a quartic potential $n=4$ (panel b) and when using the explicit full
form of the dissipation coefficient as derived from the microscopic
physics, Eq.~(\ref{dissip}), also for the quartic potential (panel
c). We see from the results shown in {}Fig.~\ref{fig1}(a) that
$\epsilon_V$ evolves faster than $Q$, thus ending inflation
eventually.  In {}Fig.~\ref{fig1}(b), we show the evolution in the
case of a quartic potential, $n=4$, where we see an opposite behavior,
$\epsilon_V$ evolves slower than $Q$, thus never ending inflation.
The behavior shown in {}Fig.~\ref{fig1} panels (a) and (b) is exactly
what we have anticipated from the analysis done above.  However, when
considering the correct full form of the dissipation coefficient
Eq.~(\ref{dissip}) in the case of the quartic potential, panel (c),
inflation does end.  This is because in the corresponding dynamics, as
shown e.g. in Ref.~\cite{Bastero-Gil:2019gao}, the temperature
decreases with the evolution.  Thus, eventually the condition $m_0 \ll
\alpha T$ required to give a $\Upsilon \propto 1/T$ in
Eq.~(\ref{dissip}) is no longer valid and the explicit form
(\ref{dissip}) will cause $Q$ to decrease faster as the system
evolves, which eventually helps end inflation. 

Let us now consider the more general case described in
Eq.~(\ref{cond-end-inf-gen}). {}For a monomial potential this
condition reduces to  
\begin{eqnarray}
8-2p+[(p-2)n+8+4c]\,Q>0.
\end{eqnarray}
As we are dealing with the range $-4<p<4$, the above condition  is
easily satisfied when 
\begin{eqnarray}
(p-2)n>-8-4c.
\label{cond-1}
\end{eqnarray}

We see from Eq.~(\ref{inc-Q}) that $Q$ increases when 
\begin{eqnarray}
(p-2)n>2p-4c.
\label{cond2}
\end{eqnarray}

{}For the range $-4<p<4$, the condition in Eq.~(\ref{cond-1}) would
also  encompass Eq.~(\ref{cond2}).  However, if $Q$ remains constant
or decreases with $e$-foldings, in which case inflation ends even more
easily, one would require $n\leq (2p-4c)/(p-2)$.  In
{}Fig.~\ref{fig2}(a) we give the general prediction from the above
inequalities, in the space of parameters $p,\, c$ and $n$, required
for graceful exit in WI in the context of the class of monomial
potentials Eq.~(\ref{Vpoly}). The regions for which $Q$ grows or
decreases during WI and there is graceful exit have been
identified. An empty (white) region indicates the parameter space
where there is no graceful exit.  The results shown in
{}Figs.~\ref{fig2}(b) and \ref{fig2}(c) exemplify the effect of having
an inflaton field dependence in the dissipation coefficient. It tends
to improve the available range of parameters  allowing graceful exit
in WI, but  at the same time it decreases the area available (i.e.,
models) for which the dissipation ratio $Q$ grows and increases the
one leading to a decreasing $Q$, which makes inflation ending more
easily according to the general discussion given above. 

One also notes from the above results that the type of graceful exit
we discuss in the  context of WI is mostly associated with the fact
that $Q$ can grow faster than $\epsilon_V$ as already pointed out
before. In this case, even if during inflation  we might have the
dynamics initially for $Q \ll 1$, i.e., start in the weak dissipative
regime,  it will certainly evolve towards the strong regime. As shown
in {}Fig.~\ref{fig2}, the regions for which WI does not end are
essentially those for which $Q$ is growing.  The CI dynamics would
only be recovered in the opposite regime, when $Q$ decreases during
the evolution and evolves to very small values at the end of
inflation.  In this case, the condition required for graceful exit
would be similar to those in CI.

\begin{center}
\begin{figure}[!htb]
\subfigure[ Evolution case for $p=-1$, $c=0$ and
  $n=2$. ]{\includegraphics[width=7.5cm]{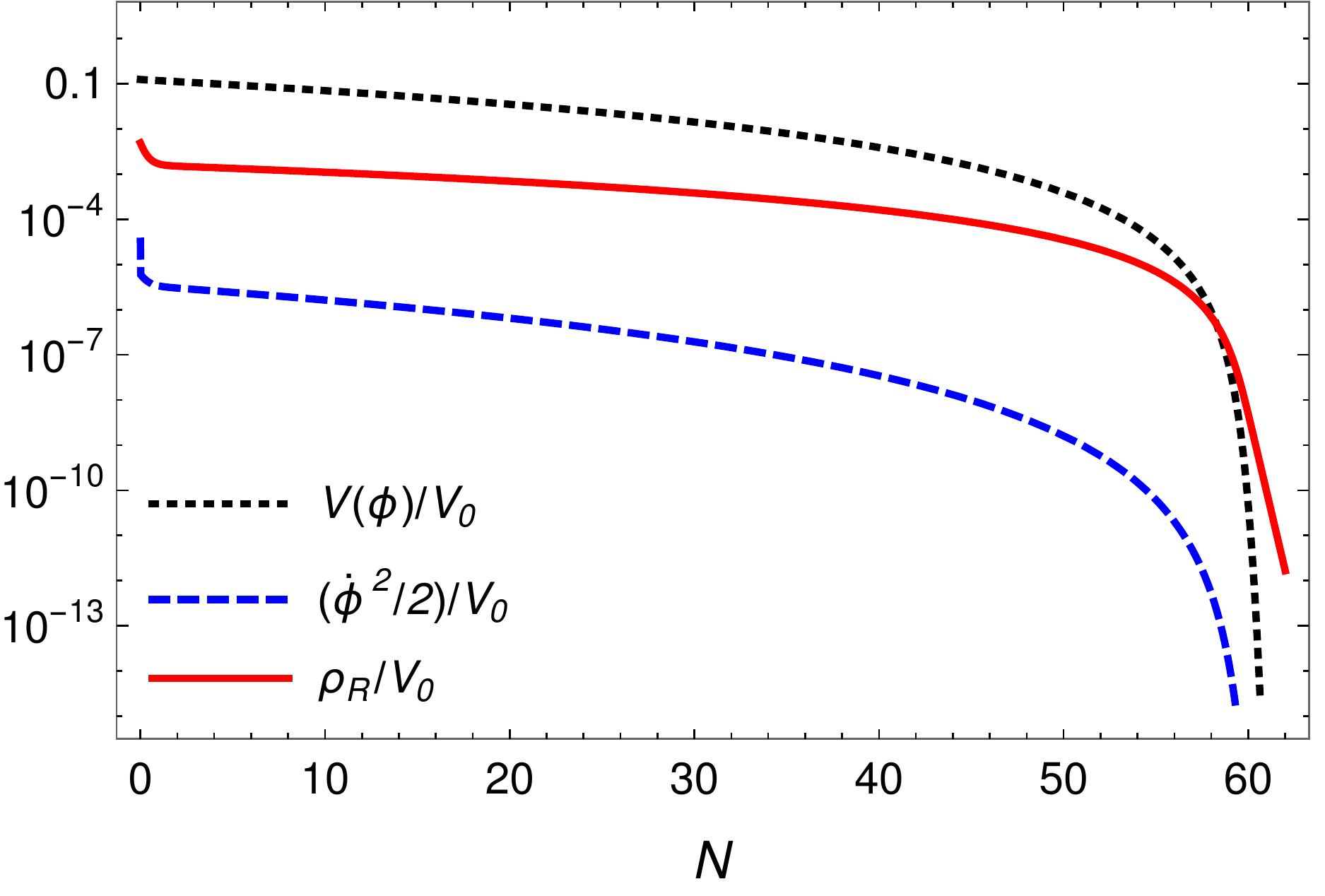}}
\subfigure[Evolution case for $p=-1$, $c=0$ and
  $n=4$.]{\includegraphics[width=7.5cm]{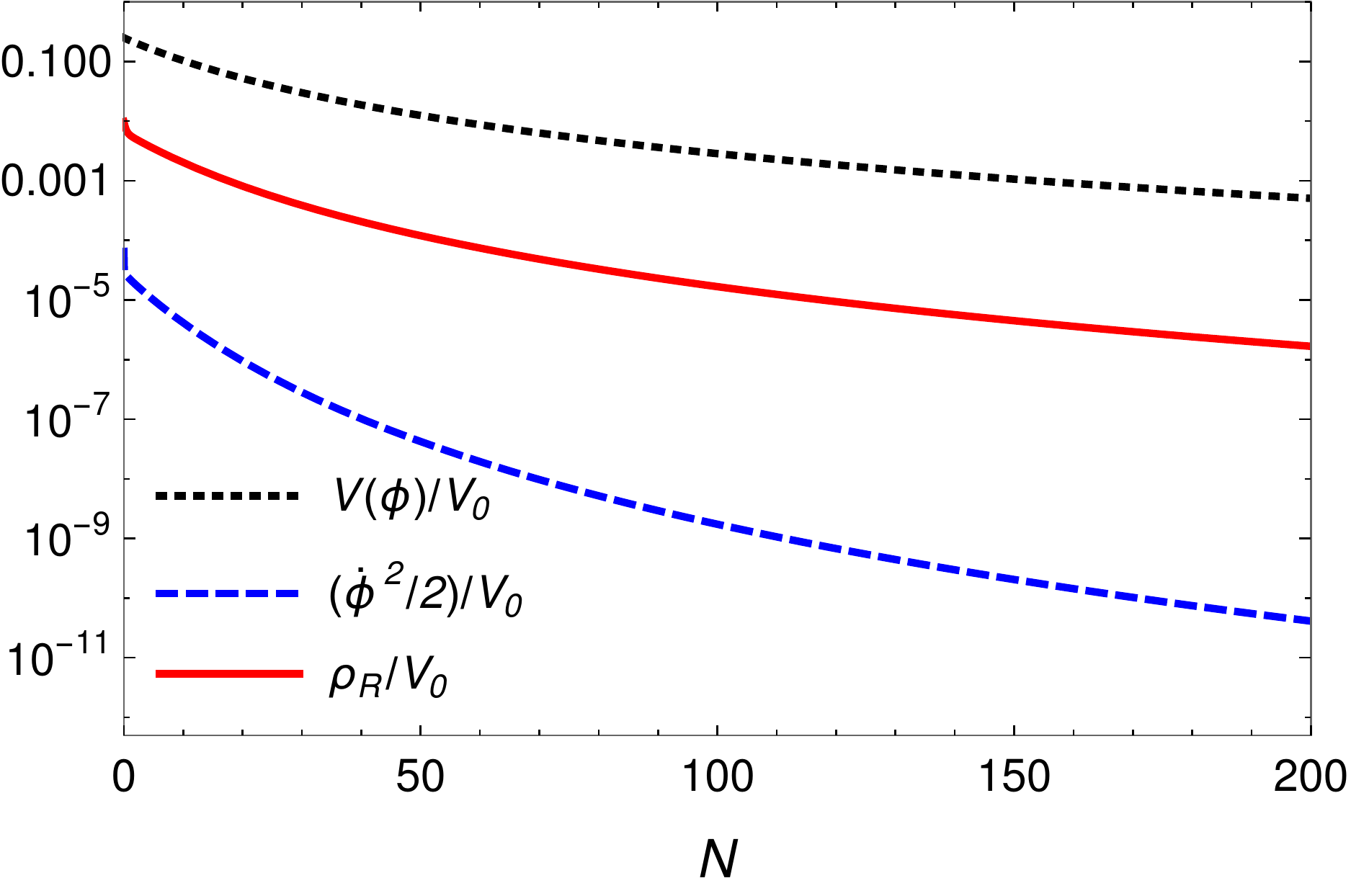}}
\caption{Example cases for the  evolution of the energy densities with
  the number of $e$-foldings $N$. Parameters used were again based on
  the ones considered in Ref.~\cite{Bastero-Gil:2019gao},
  corresponding e.g. to $C_\Upsilon\simeq 5.13$, $M\simeq 5.5 \times
  10^{-5} M_{\rm Pl}$ in Eq.~(\ref{Upsilon}),  with an initial
  dissipation ratio $Q_0 \simeq 100$.}
\label{fig3new}
\end{figure}
\end{center}

As already emphasized in the Introduction, even though there is
radiation production due to the dissipation, this does not mean that
radiation will overtake the inflaton energy density and end
inflation. This is illustrate by using again the example shown in
{}Fig.~\ref{fig1} for the cases in panels (a) and (b). In
{}Fig.~\ref{fig3new} we show the evolution for the energy densities in
the cases of an inversely proportional in the temperature dissipation
coefficient for the cases of a quadratic (panel a) and quartic (panel
b) inflaton potentials. These correspond to the cases where inflation
can end in the former, which lies inside the red region shown in
{}Fig.~\ref{fig2}(b),  and never end in the latter, which lies inside
the white region shown in {}Fig.~\ref{fig2}(b).  Note that in the
first case radiation overtakes the inflaton energy density ending
inflation smoothly in a radiation dominated regime. In the second case
there is always a nonvanishing radiation energy density being
produced, yet it gets damped in a rate faster than the decrease in the
inflaton energy density and inflation does not end. This behavior
persists for much longer times (efoldings) than the  ones shown in the
figure.

\subsubsection{Negative $\eta_V$}

\begin{center}
\begin{figure*}[!htb]
\subfigure[3d parameter space
  $(c,p,n)$]{\includegraphics[width=5.7cm]{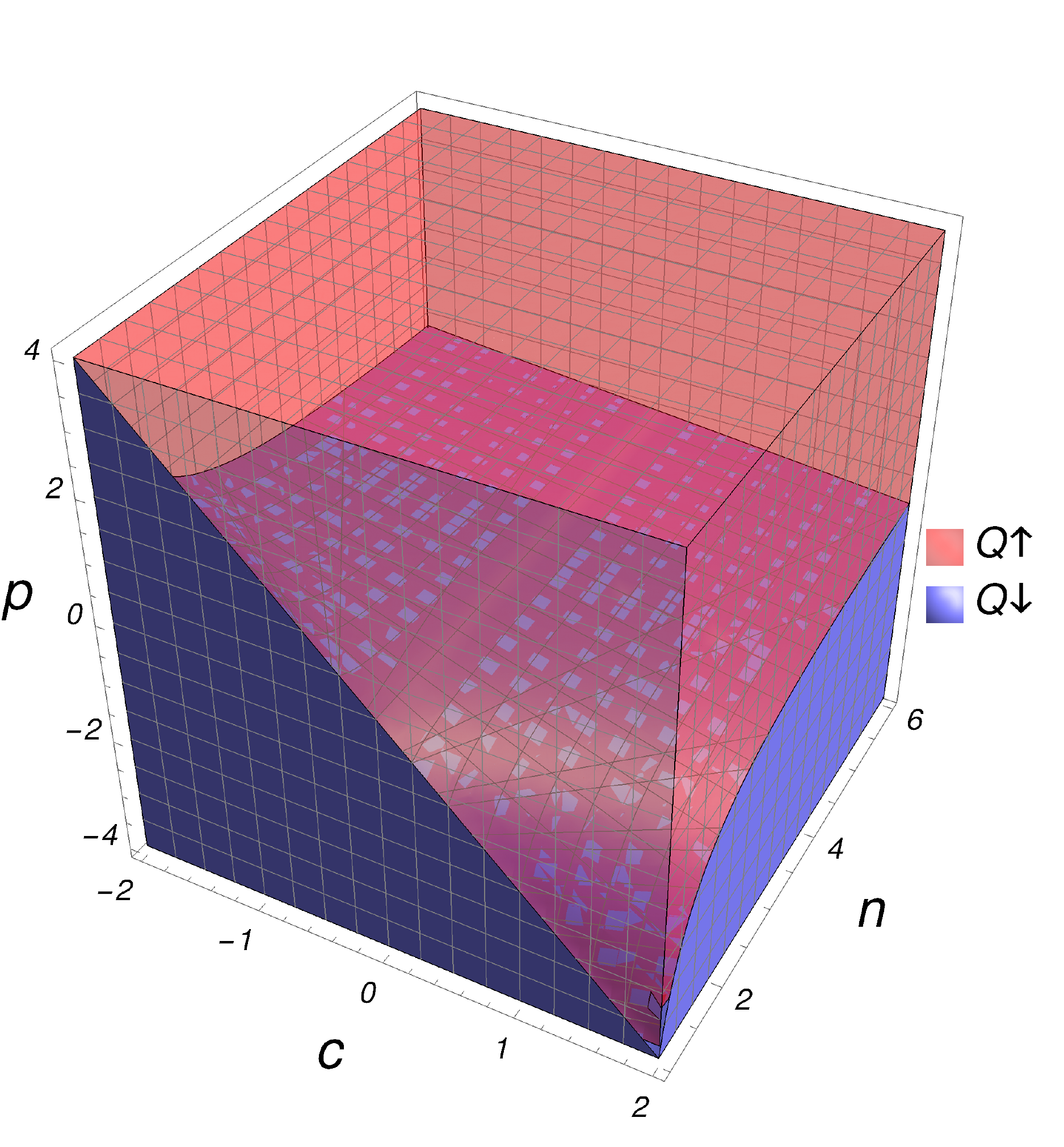}}
\subfigure[Plane $c=0$]{\includegraphics[width=5.2cm]{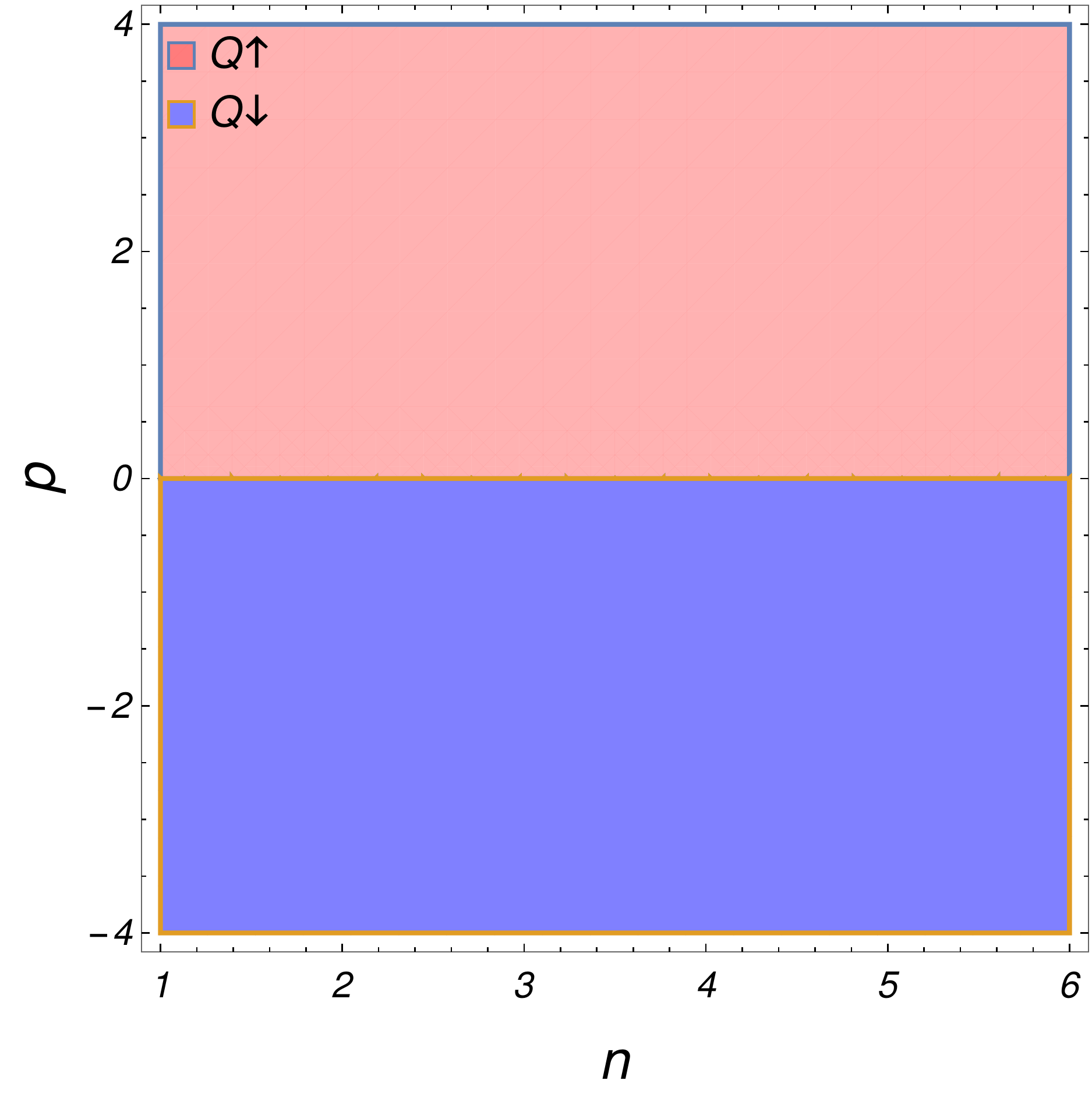}}
\subfigure[Plane $c=2$]{\includegraphics[width=5.2cm]{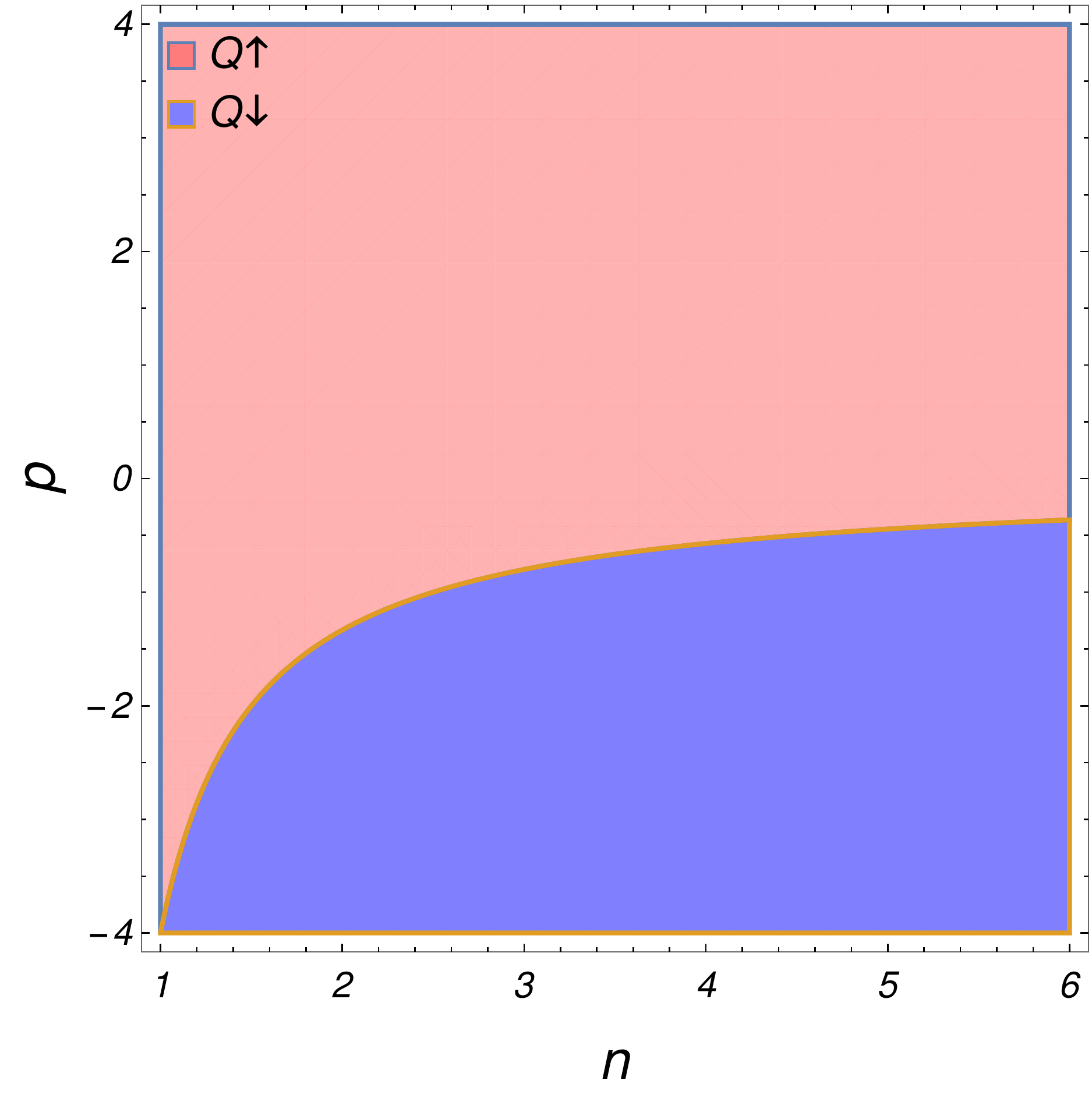}}
\caption{The  parameter space for graceful exit in the case of the
  hilltop class of  inflaton potentials Eq.~(\ref{VHiggs}). The shaded
  areas indicate where $Q$ grows or decreases and WI has a graceful
  exit.}
\label{fig3}
\end{figure*}
\end{center}

Next, we consider the case for concave inflaton potentials for which
$\eta_V$ is negative. In such a case, the condition for graceful exit
becomes
\begin{eqnarray}
&&\epsilon_V \left[ 2(4-p) + (6+p)Q \right] + | \eta_V |
  \left(4-p+4Q\right)  \nonumber\\ &&+ 2 c\, Q\, \kappa_V >0.
\label{concave}
\end{eqnarray}
As an example, we can consider the hilltop-like class of potentials
for the inflaton given by
\begin{equation}
V(\phi) = V_0 \left[ 1 - \left( \frac{\phi}{\phi_0} \right)^{2n}
  \right]^2,
\label{VHiggs}
\end{equation}
with $n \geq 1$ and with inflation taking place around the top
(plateau) of the potential, $|\phi| \ll \phi_0$.  We also consider
that $\phi_0$ is sufficiently large such that inflation ends before
the inflection point of the potential, thus, inflation takes place
exactly in the concave part of the potential.  The condition for
graceful exit in this case then becomes $8n >2c+4$ in the regime of
strong dissipation, while for weak dissipation we have that $8n >
p(2n-1)+4$. In {}Fig.~\ref{fig3}(a) we show the corresponding  region
for graceful exit, along also with the regions where we have a growing
or decreasing $Q$.  The  {}Figs.~\ref{fig3}(b) and  \ref{fig3}(c) show
the $c=0$ and $c=2$ planes, like what we have shown in the previous
case. Notice that here all the parameter space allows for graceful
exit.

Even though we have used in the above study the potential
Eq.~(\ref{VHiggs}), it is easy to check that the results also apply to
other classes of concave potentials. Thus, the concave potentials do
lead generically to graceful exit in a WI setup. We also note that
apart from WI taking place in strong dissipative regimes, a growing
$\epsilon_V$ is sufficient to ensure graceful exit here, just as in
cold inflation. 

\subsection{Constant $\epsilon_V$}
\label{sec3B}

Let us now consider the case of a constant $\epsilon_V$. Apart from
the cosmological constant $\Lambda$, the classic example of such a
potential is the runaway potential or the exponential potential,
\begin{eqnarray}
V(\phi)=V_0e^{-\alpha\phi/M_{\rm Pl}},
\label{runaway}
\end{eqnarray}
which leads to constant slow-roll parameters
$\eta_V=2\epsilon_V=\alpha^2$. Such a potential drives a  power-law
type inflation in cold inflation and does not lead to graceful
exit~\cite{Liddle:1988tb} when $\alpha < \sqrt{2}$, while for $\alpha
\geq \sqrt{2}$ there is not even an accelerated expansion. 

However, if $Q$ decreases in a WI inflation setup, then there is a
possibility to have both inflation and graceful exit in such
inflationary scenarios when $\alpha > \sqrt{2}$. Accelerated expansion
happens because we can arrange for $\epsilon_V = \alpha^2/2 < 1+Q$ for
sufficiently large dissipation ratio $Q$. And graceful exit happens
when $Q$ decreases. It is easy to see from Eq.~(\ref{q-evo}) that $Q$
decreases when $p>2$ and it is independent of the value of $c$
whenever $\alpha\phi/M_{\rm Pl} > 1$.  This phenomenon has already
been noted in Refs.~\cite{Lima:2019yyv, Goswami:2019ehb}.  It is to
note that in this case, the inflaton will keep on slow-rolling if $Q$
keeps on increasing, and there will not be no graceful exit as such a
runaway potential possesses no minimum (just as it happens in the case
of cold inflation discussed above).

\subsection{Decreasing $\epsilon_V$}
\label{sec3C}

Equation~(\ref{epsilon-evo}) tells us that $\epsilon_V$ decreases
whenever 
\begin{eqnarray}
\frac{\epsilon_V}{\eta_V}<\frac12.
\end{eqnarray}
Clearly, such potentials are not employed in cold inflation as they
will not lead to graceful exit. However, as both $\epsilon_V$ and $Q$
can decrease simultaneously in WI, there is a sliver of chance that
inflation might end in such scenarios.  One such potential of the form 
\begin{eqnarray}
V(\phi)=V_0\left(1+e^{-\alpha\phi/M_{\rm Pl}}\right),
\label{pot-dec-eps}
\end{eqnarray}
has been employed in the MWI model
recently~\cite{Goswami:2019ehb}. However, it is to note that WI takes
place when $\epsilon_V<1+Q$. Then, in such decreasing $\epsilon_V$
scenario, one needs to start with $\epsilon_V>1$ and $Q$ not only
requires to fall faster than $\epsilon_V$ but it must fall much faster
such that $1+Q$ crosses $\epsilon_V$ before $\epsilon_V$ decreases
below unity. {}Fulfilling such a condition is rather challenging and
inflation would not end in such scenarios in most of the cases. 

\begin{figure}
\begin{center}
  \includegraphics[width=7.5cm]{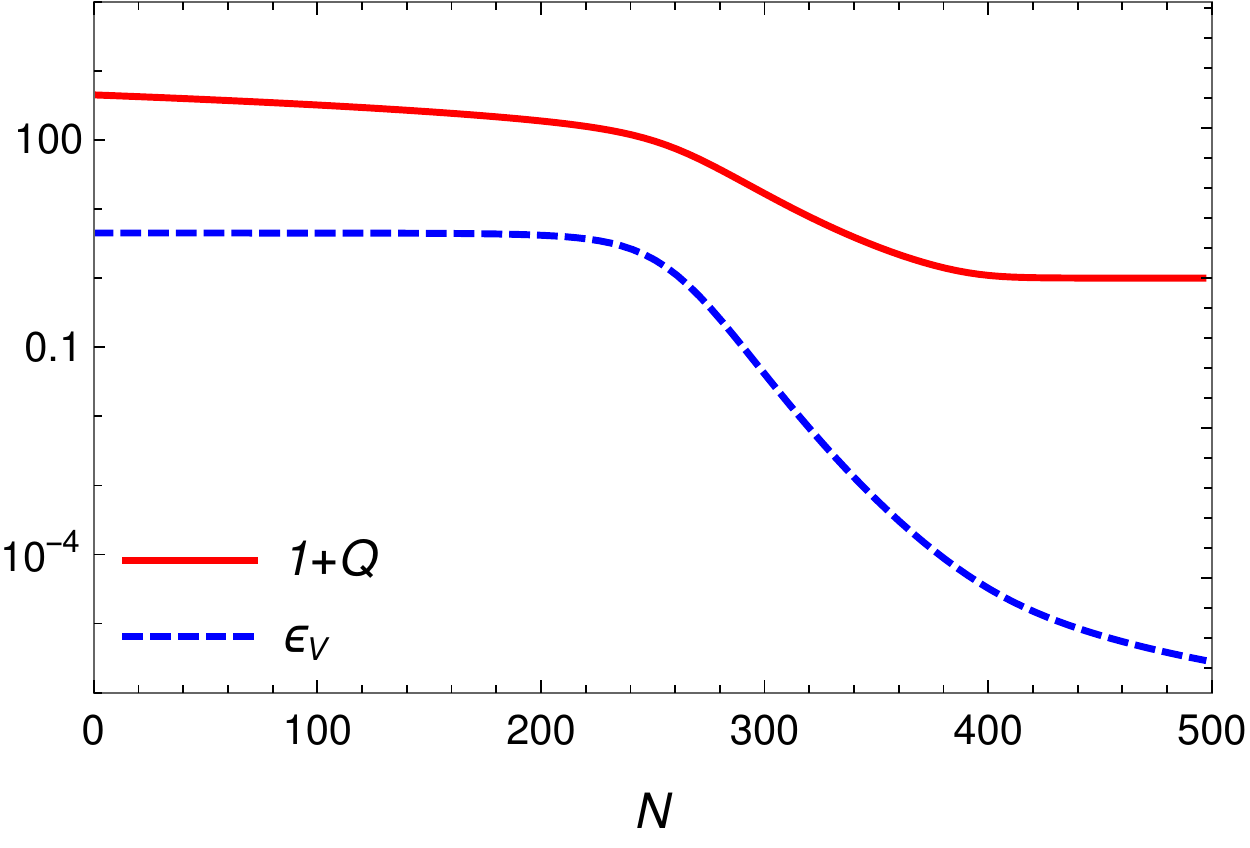}
  \caption
 {Typical behavior of $Q$ and $\epsilon_V$ in the MWI model and with a
   form a potential given in Eq.~(\ref{pot-dec-eps}).  }
  \label{Tcube-dec-eps}
  \end{center}
\end{figure}

We have evolved the full background equations with the above potential
applied in the MWI model (e.g., $p=3,c=0$), and have showed in
{}Fig.~\ref{Tcube-dec-eps}  the typical behavior of $\epsilon_V$ and
$Q$ in such scenarios.  As shown in the example of
{}Fig.~\ref{Tcube-dec-eps},  even though $Q$ starts to decrease faster
than $\epsilon_V$ initially, it does not fall faster enough.
$\epsilon_V$ eventually drops below unity and $1+Q$ saturates to 1,
and, hence, inflation is not terminated.


\section{Further discussions and possible applications}
\label{applications}

Let us discuss some possible applications of the results obtained in
this paper. 

We should recall that the primordial spectrum of scalar curvature
perturbations is strongly affected by the dissipation in
WI~\cite{Graham:2009bf,BasteroGil:2011xd,Ramos:2013nsa,Bastero-Gil:2014jsa,Li:2018wno}.
In particular, a dynamical regime for which $Q$ grows invariably leads
to a growing spectral tilt $n_s$, thus to a bluer spectrum. In the
opposite regime, for which we have a decreasing $Q$, the effect has
been shown to produce a decreasing $n_s$, i.e., a redder spectrum (for
systematic analysis of the effect of the behavior of the dissipation
on the power spectrum, see, e.g.,
Refs.~\cite{Graham:2009bf,BasteroGil:2011xd,Bastero-Gil:2014jsa}).  To
see these effects explicitly on the observables, let us recall that in
WI the primordial curvature power spectrum is of the general
form~\cite{Ramos:2013nsa,Benetti:2016jhf}
\begin{equation} 
\Delta_{{\cal R}}(k) =  \left(\frac{ H^2}{2 \pi\dot{\phi}}\right)^2
      {\cal F} (Q),
  \label{Pk}
\end{equation}
where all quantities are meant to be evaluated at the Hubble radius
crossing, $k = aH$.  The function ${\cal F} (Q)$ in Eq.~(\ref{Pk})
strongly depends on the form of the dissipation coefficient and the
consequent dynamics (see, e.g.,
Refs.~\cite{Graham:2009bf,BasteroGil:2011xd,Bastero-Gil:2014jsa}).
{}From Eq.~(\ref{Pk}), the spectral tilt $n_s$ is defined as
\begin{eqnarray}
n_s-1 &=& \frac{d \ln \Delta_{{\cal R}}(k)}{d \ln k} \simeq \frac{d
  \ln \Delta_{{\cal R}}(k)}{d N} \nonumber \\ & \simeq & \frac{-6
  \epsilon_V + 2 \eta_V + 2 \frac{dQ}{dN} }{1+Q} +  \frac{dQ}{dN}
\frac{d \ln  {\cal F} (Q)}{dQ}.
\label{ns}
\end{eqnarray}
Typically, the last term dominates in many of the WI
realizations. Thus, for WI models for which $Q$ grows with the number
of efolds but have a decreasing  ${\cal F} (Q)$ with $Q$, like e.g.,
in the models with a dissipation coefficient with negative powers in
the  temperature~\cite{Motaharfar:2018zyb}, they will lead to a redder
scalar power spectrum.  Other models leading also to a growing $Q$
with the number of efolds but an  increasing ${\cal F} (Q)$ with $Q$
will lead to a bluer spectrum. Similarly, a decreasing $Q$ with $N$
will have an effect on the observables. Hence, we see that the
dynamical behavior of the dissipation ratio is fundamental to
determine whether we can have a redder or bluer spectrum.  Thus,
determining possible regimes for a growing or decreasing dissipation
ratio can be of fundamental relevance for WI model building. {}For
instance, models that in the cold inflation context are excluded
because the spectrum is too red or it is too blue, can be rendered
observationally consistent in the WI picture by choosing models
leading to either a growing or to a decreasing dissipation ratio,
respectively.

It is important to also point out that the issue of graceful exit in
WI may not be a negative feature but in some cases might be a
desirable outcome. A sustained acceleration regime in the late
Universe, attributed to dark energy, might conceivably  emerge as a
special case of an overdamped dynamical regime for a scalar field.  In
this case, given a potential with a minimum at $\phi=\phi_0$ and if
the approach to $\phi_0$ is asymptotic, as discussed above in the
context of the graceful exit problem in WI, the scalar field can
remain close to the minimum $\phi_0$, then with $V(\phi\neq \phi_0)>0$
the evolution would mimic closely that of a cosmological
constant. This is particularly important when the energy density
varies very slowly. {}For instance, from the slow-roll
equations~(\ref{slowrolleqs}), we can deduce that the rate of change
of the energy density of the scalar field in a Hubble time can be
expressed as~\cite{Bartrum:2014fla}
\begin{equation}
\frac{1}{H} \left|\frac{\dot\rho_\phi}{\rho_\phi} \right|  \simeq 2
\frac{\rho_\phi}{\rho_{\rm total}} \epsilon_{wi} \ll 1,
\label{de}
\end{equation}
where we have also used Eq.~(\ref{epsH}) and $\rho_{\rm total}$
denotes the total  energy density at that given time in the late-time
Universe, where we also expect that $\rho_\phi \ll \rho_{\rm total}$,
if the energy density of the scalar field is not the dominant energy
component.  The result given by Eq.~(\ref{de}) shows that whenever
$\epsilon_{wi} \ll 1$, which happens in particular when the conditions
for not having graceful exit and that we have discussed are met,  then
the scalar field dissipates very little of its energy density on
cosmological time scales.  Thus, the overdamped regime that can
potentially lead to a graceful exit problem in the early Universe, at
late times can also potentially sustain a (slowly-varying)
cosmological  vacuum energy term mimicking a cosmological constant.
Thus, the result we have obtained in the present paper can certainly
be used as a guide  in the development of possible quintessential-like
scalar field models where the late-time  accelerated expansion might
eventually emerge as a consequence of dissipative effects (see, e.g.,
Ref.~\cite{Lima:2019yyv} for a specific attempt in this context).

\section{Conclusions}
\label{conclusions}

To conclude, we have shown that the process of graceful exit in WI is
a more rich phenomenon than it is in cold inflation. Graceful exit in
WI depends on three independent choices: The form of the potential
(hence the behavior of $\epsilon_V$ during inflation), the choice of
the WI model (hence the form of the dissipative coefficient) and
whether we want WI to take place in a weak or in a strong dissipative
regime. However, we have shown that there are cases, where a potential
with growing $\epsilon_V$ in time (or with $e$-foldings) is good
enough to exit WI gracefully, as it happens in a cold paradigm. This
happens in the cases of inflation happening in the plateau region of
the concave potentials or in cases where $Q$ is decreasing with
$e$-foldings.  In the rest of the cases, one needs to make sure that
$\epsilon_V$ grows faster than $Q$ in order to exit gracefully.  We
have also shown that WI is capable of terminating inflation even in
cases where cold inflation fails to exit gracefully, such as in the
case of runaway potentials.

It is to note that the reason behind no graceful exit in these cases
of WI differs from the usual reason of getting trapped in a false
vacuum, as it happens in cold inflation. Mostly, $Q$ increasing faster
than $\epsilon_V$ (where $\epsilon_V$ evolves) or a simply growing $Q$
in cases of constant $\epsilon_V$ leads to no graceful exit in WI. It
is to recall that the parameter $Q$ incorporates extra frictional term
in the equation of motion of the inflaton field, and an increasing
$Q$, in these cases, will make such an equation overdamped. Therefore
inflation can only end in an asymptotic time, resulting in a realizing
of never-ending inflation, and hence no graceful exit.

The situation is particularly delicate in the case of chaotic like
monomial potentials. In this case, not all parameter space allows
graceful exit, as far as the simple parametrization for the
dissipation coefficient, given by Eq.~(\ref{Upsilon}), is concerned.
We have demonstrated the issues involved in this case in an example
making use of a dissipation coefficient  $\Upsilon \propto 1/T$. In
this case, the simple parametrization Eq.~(\ref{Upsilon}) is not
enough to get the complete dynamics in WI and for studying graceful
exit. We have also to consider the microscopic physics leading to this
coefficient and the consistency conditions leading to the simple form
for the dissipation coefficient.  Given that most phenomenological
studies involving WI make use of the simple parametrization given by
Eq.~(\ref{Upsilon}), these results prompt to a word of caution in the
studies of the dynamics in those models if details of the possible
microscopic physics are left unchecked.  Likewise, we have also shown
that whether or not the dissipation coefficient depends on the
inflaton amplitude can change the conclusions regarding graceful exit
in a significant way.
 
Our study that we have performed in this paper have also identified
regimes for which the dissipation ratio $Q$ can either grow or
decrease with the number of efolds, depending on the different
combinations of dissipation coefficient and primordial potentials.
This is particularly important in the context of obtaining
observational predictions in WI.  It would also be interesting to
carry out a similar analysis as the one performed here to other
dynamics involving WI, like in noncanonical type of
models~\cite{Li:2018riw,Zhang:2019bgv,Zhang:2020sbk}.

\section*{Acknowledgments}

S.D. would like to thank R. Rangarajan for providing useful
references.  The work of S.D. is supported by Department of Science
and Technology,  Government of India under the Grant Agreement number
IFA13-PH-77 (INSPIRE Faculty Award).  R.O.R. is partially supported by
research grants from Conselho Nacional de Desenvolvimento
Cient\'{\i}fico e Tecnol\'ogico (CNPq), Grant No. 302545/2017-4, and
Funda\c{c}\~ao Carlos Chagas Filho de Amparo \`a Pesquisa do Estado do
Rio de Janeiro (FAPERJ), Grant No. E-26/202.892/2017.
 

\end{document}